\documentstyle[preprint,aps,psfig]{revtex}
\begin{document}
\draft

\title{ LOOP VARIABLES FOR COMPACT TWO-DIMENSIONAL \\
       QUANTUM ELECTRODYNAMICS}

\author{Rodolfo Gambini$^{\dagger}$, Hugo A.
Morales--T\'ecotl$^{*,\S}$,\\
Luis F. Urrutia$^{\ddagger}$ and J. David Vergara$^{\ddagger}$.\\
$^{\dagger}$Departamento de F\'{\i}sica, Facultad de Ciencias,\\
Universidad de La Rep\'ublica,  \\
Trist\'an Narvaja 1674, Montevideo. \\
$^*$Departamento de F\'{\i}sica,
Universidad Aut\'onoma Metropolitana-Iztapalapa,\\
Apartado Postal 55-534, 09340, M\'exico D.F.\\
$^{\S}$International Centre for Theoretical Physics,\\
P.O. Box 586, 34100 Trieste.\\
$^{\ddagger}$Instituto de Ciencias Nucleares, Universidad Nacional
Aut\'onoma de M\'exico, \\
Apartado Postal  70-543, 04510,  M\'exico D.F.
}

\maketitle
\tightenlines 
\begin{abstract}
Variables parametrized by closed and open curves are defined to
reformulate compact U(1) Quantum
Electrodynamics in the circle with a massless fermion field. It is found
that the gauge invariant nature of these variables accommodates into
a regularization
scheme for the Hamiltonian and current operators that is specially well
suited
for the study of the compact case. The zero mode energy spectrum, the
value of the axial anomaly  and the anomalous commutators
this model
presents
are hence determined in a manifestly gauge invariant manner. Contrary to
the non compact case, the zero mode spectrum is not equally spaced and
 consequently the theory does not lead to the spectrum of a free scalar
boson.  All the states  are invariant under large gauge transformations.
In particular, that is the case for the vacuum, and consequently the
$\theta$-dependence does not appear.
\end{abstract}
\maketitle
\section{Introduction}
\baselineskip=12pt

Back in 1962 Schwinger considered the possibility that a vector
gauge field can
imply a nonzero mass gauge particle. He showed this is indeed
the case for
an exactly solvable model,
namely two-dimensional Quantum Electrodynamics (QED$_2$) with
a massless fermion field
\cite{Schwinger}.
This simple, although non-trivial, field theory has become since
an arena to probe
different  aspects
of quantum field theory \cite{general,ADAMS}. For instance,
the axial
anomaly, the charge screening and the bosonization presented by the  
model,
together with alternative methods of solution,
have been studied by different authors
\cite{Manton,HetHo,Shifman,Link,IsoMurayama,HallinLiljenberg}. Of  
particular
interest are
the results from
QED$_2$ that might shed light on the non-perturbative
features not only of QED$_4$ but also of QCD$_4$  and
quantum gravity in its gauge-theory-like formulation
\cite{G-P,RovelliSmolin,Gambini}.

Being QED$_4$ the gauge theory {\em par excellence} it was
considered worth adopting
loop variables techniques in its description
\cite{GambiniTrias,RovelliSmolin}
because
in this way gauge invariance could be explicitly implemented.
Lattice QED$_4$ has been successfully developed along these lines
and computational calculations have been improved
with respect to simulations \cite{FortGambini}.
Also, a  four-dimensional (continuum) gravitational analogue
constructed out of  gravity plus  fermions
has been studied in
this framework \cite{HugoRovelli}. In every case the presence of
fermions is
automatically accounted for by including open curves, besides loops,  to
parametrize operators and state vectors. The fermions necessarily
stand at the
end points of the open curves.
Moreover,  the dynamics gets
geometrically coded
into the  breaking, rejoining and rerouting of such open curves and
loops through
their
intersection
points. This is deeply significant for non-perturbative quantum gravity
where the role of diffeomorphism invariance is central and such a
description
naturally fits in \cite{HugoRovelli}. For the QED$_4$  case
this geometrical picture of dynamics led to useful criteria to
approximate the
strong coupling regime \cite{FortGambini}.

Wilson loops are computed in terms of the holonomy elements associated
to the parallel transport along closed curves. In the Maxwell case the
holonomies are elements of the U(1) group. They are not only invariant
under small gauge transformations generated by the Gauss law constraint,
but also under large gauge transformations. That is the reason why they
naturally describe compact electrodynamics, which is also  characterized by the
property that the range of  $A_1(x)$, in the Weyl gauge for the two-dimensional case, is the circle
instead of the real line as it is in  the standard noncompact case. This fact is well known and
has far reaching consequences in higher dimensions. Polyakov  \cite{Po}
has argued that it is necessary to decide, based on physical
grounds, what
version of QED is realized in nature. In particular,  the fact that
in the
non abelian case the non compact version cannot be formulated on a
discrete lattice leads him to consider that, if QED arises as a
subgroup of
some nonabelian gauge theory,  we are necessarily dealing with the
compact
version.  Earlier discussions of compact QED in the lattice can be found in Refs. \cite{COMPACT}, for
example. The Schwinger model has been recently studied in the hamiltonian
\cite{FA} and lagrangean\cite{F} lattice loop representation. In these
papers, it is shown that the chiral symmetry is broken and the
$\theta$-dependence of the vacuum is not present.   

In what concerns the case of higher dimensions, it has been shown that
monopoles arising in the compact abelian sector  of 2+1 QCD play a
fundamental role in the confinement process. In Polyakov's analysis
loops are crucial to understand this process. In particular,  he has
recently shown that the loop world sheets acquires string like degrees
of freedom due to the presence of a diluted gas of monopoles
\cite{Poly96}.

In this work, loop variables are
introduced for the compact Schwinger's model along the
lines of \cite{FortGambini,HugoRovelli}.
Starting with the
canonical analysis of QED$_2$,  which yields the Gauss law first-class
constraint, loop variables are defined such that they have zero Poisson
brackets with it. They form a closed algebra and turn out to be enough to
describe the dynamics.
The Hamiltonian is reexpressed as a limit
of some of these loop variables when the curves shrink down to a point.
The quantum theory is defined such that the Poisson
algebra becomes a commutator algebra and the loop representation is built
by choosing one of these loop operators to create a state with an extra
loop (open curve) out of an arbitrary state and then using the operator
algebra.
Thus, it is possible to work entirely in the loop representation.
In order
to recover  the standard (local) physical information, like  the energy
spectrum  for example,  one has
to take the corresponding limit of the loops (open curves)
shrinking to  a point.
Before determining the properties of the energy spectrum of the
full theory,
we found it convenient to study the semiclassical situation where
there is
a quantum fermion field
interacting with a classical electromagnetic field. Some technicalities
become more transparent if use is made of  a bi-local  Fourier transform
of the operators parametrized by open curves. Details are given in the
appendix.
This procedure yields a built-in {\em gauge-invariant} point-split
regularization for the fermion vacuum energy. In this semiclassical
context,
the vacuum to vacuum expectation value for the divergence of the
axial current
produces the known value for the axial anomaly in a straightforward
manner.  Also, the vacuum expectation value of the so called
anomalous commutators are directly derived from the loop variables
algebra.
Remarkably
enough, the corresponding Poisson brackets algebra already contains the
relevant information.  Using the separability of the full Schroedinger
equation for the system,
the zero mode sector of  the spectrum is considered
next. The non-equally spaced results  for the zero mode energy found in this case seem to  
indicate that,
in the general compact case, the spectrum does not correspond to a
free massive boson. By considering the limit in which
the length of the $S^1$ spatial slice goes to zero one recovers the 
typical harmonic oscillator spectrum
for the zero mass mode together with its free bosonic behavior, which is  characteristic of the
non-compact Schwinger Model. 

As we have previously emphasized, loop variables naturally describe  
 a compact
version of the electromagnetic interaction.
In order to recover  the noncompact  theory it is necessary to introduce
additional
angular variables, which are conjugated to the integer numbers   
characterizing
the large
gauge transformations. A detailed discussion of this important  
issue can be
found in Ref. \cite{GAMB5}.

The organization of the paper is as follows.
In section II we translate the local classical dynamics of QED$_2$ into
loop variables.
The corresponding algebra is displayed there. The quantum counterpart is
then exhibited in section III,  where state functionals are loop/curve
parametrized; hence defining the loop representation. In section IV, the
analysis is carried out taking the fermion field as a quantum entity
evolving in
the external electromagnetic field. Hereby the fermion vacuum is
found. In section V, the
well known chiral anomaly coming from the non-conservation of the axial
current is computed. Section VI contains the calculation of the anomalous
commutators.
The zero mode sector of the theory is finally analyzed in section
VII based on the external electromagnetic field approach of the
previous section.
Finally, section VIII contains some general remarks on the
loop approach for QED$_2$
together with possible future developments. The appendix presents
the explicit relation between loop operators and their useful
bi-local Fourier transforms, making more transparent the analysis here
presented.

\section{Classical framework}

Our starting point is the {\em real} Lagrangian density
\begin{equation}
{\cal L} = -\frac{1}{4} F_{\mu\nu}F^{\mu\nu}
           +\frac{\hbar}{2}\bar{\psi} \gamma^{\mu} \left(i\partial_{\mu}
           - e A_{\mu}\right)\psi
           -\frac{\hbar}{2}
           \left[\left(i\partial_{\mu} + e
A_{\mu}\right)\bar{\psi}\right]
           \gamma^{\mu} \psi
\label{LAG}
\end{equation}
where $F_{\mu\nu}=\partial_{\mu}A_{\nu} -\partial_{\nu}A_{\mu}$
and $\bar{\psi}=\psi^\dagger \gamma^0$ is a  Grassmann valued
fermionic field.
Since space here is $S^1$, we will require periodic(antiperiodic)
boundary conditions for the fields
\begin{equation}
A_{\mu}(x+L) = A_{\mu}(x), \quad
\psi(x+L) = -\psi (x),
\label{BC}
\end{equation}
where  $L=2\pi r$ is  the length of the circle.
The gamma matrices are:
$\gamma^0=\sigma_1,\; \gamma^1=-i\sigma_2,\; \gamma^5=\gamma^0\gamma^1=
\sigma_3$, where $\sigma_i$ are the standard Pauli matrices.
We use the  signature $(+,-),\;\; i.e.\;\; \eta_{00}=-\eta_{11}=1$.

The Lagrangian density (\ref{LAG}) is invariant under the
following gauge
transformations
\begin{equation} \psi \rightarrow  e^{ie \alpha(x,t)} \psi,
\quad A_\mu \rightarrow A_\mu - \partial_\mu \alpha (x,t). \label{GGT}
\end{equation}
There are two families of gauge transformations: (i) those continuously
connected to the identity,
called small gauge transformations, characterized by  the  function
$\alpha = b(t) e^{i 2 \pi n x  / L} $ which is periodic in $x$ and
preserves
the boundary conditions
(\ref{BC}). The second family corresponds to the so called large gauge
transformations, which is determined by the
 non-periodic functions $\alpha={2 \pi
n \over e L}x , \ n=\pm 1, \pm 2, \dots $.
The boundary conditions (\ref{BC}) are also preserved in this case.

After the standard canonical analysis the Hamiltonian density
becomes
\begin{equation}
{\cal H} = \frac{1}{2} E^2
           - \frac{i\hbar}{2} \psi^\dagger\sigma_3
           \left(\partial_1 + i e A\right)\psi
   + \frac{i\hbar}{2} \left[\left(\partial_1-i e A\right)\psi^*\right]
           \sigma_3 \psi
           - A_0\,{\cal G}.
\label{HAM}
\end{equation}
Here $E=F_{01}$, $A=A_1$ and
\begin{equation}\label{GL}
{\cal G}=\partial_1 E - e\hbar \psi^*\psi
\end{equation}
is the Gauss law constraint.
The  boundary term associated with the integration of  $\partial_x(EA_0)$
yields no contribution
to the Hamiltonian  because of the boundary conditions on the
vector potential $A_{\mu}$.
In what follows $\psi = (\psi_1,\psi_2)^\top$,
$\top$ denoting transposition. The charge density is
given by $\rho(x)= e (\psi_1^*\psi_1 + \psi_2^*\psi_2$).
We are working in units such that $c=1$, $\hbar  \neq 1$ and we take
mass $[g]$ and length $[cm]$ as the basic ones. In this way the
corresponding
dimensions are: $ \hbar = [g \  cm],\  E=[\sqrt{g\over cm}],\
A=[ \sqrt{g \  cm}], \
eA=[{1\over cm}], \  \psi_1^* \psi_1=[{1\over cm}],
\ \psi_2^* \psi_2=[{1\over cm}]$ and $   \hbar e^2 L^2 $ is a
dimensionless
quantity. Both the combinations ${\hbar \over L }$ and  $ e
\hbar^{3\over2}$
have the dimensions of mass.

The resulting Poisson brackets algebra at equal times  is
\begin{eqnarray}
\left\{ A(x), E(y) \right\} &=& \delta(x,y) \nonumber \\
\left\{ \psi_{\alpha}(x), \psi^*_{\beta}(y) \right\}
      &=& -\frac{i}{\hbar} \delta_{\alpha\beta}
          \delta(x,y)\;\;\;\;\;\alpha=1,2\,.
\label{PB}
\end{eqnarray}

\subsection{Loop variables}
In order to take into account  the Gauss law  {\it ab initio},
one can adopt the following
gauge invariant non local variables
\cite{FortGambini}:
\begin{eqnarray}
T^0(\gamma) &=& \exp\{ie\oint_{\gamma}dx\,A(x)\}, \label{T0}  \\
\Pi_0(\eta_x{}^y) &=& \psi_1^*(x) U(\eta_x{}^y) \psi_2(y), \label{PI)} \\
\Pi_1(\eta_x{}^y) &=& \psi_1^*(x) U(\eta_x{}^y) \psi_1(y),
\;\label{PI1} \\
\Pi_2(\eta_x{}^y) &=& \psi_2^*(x) U(\eta_x{}^y) \psi_2(y), \label{PI2} \\
\Pi_3(\eta_x{}^y) &=& \psi_2^*(x) U(\eta_x{}^y) \psi_1(y), \label{PI3}
\end{eqnarray}
with $U(\eta_x{}^y)=\exp\{ie\int_{\eta_x{}^y}dz\,A(z)\}$
and, of course, $E(x)$ which is gauge invariant by construction.
The open paths ${\eta}_x^y$ are always arcs of circumference starting
at the point $x$ and ending at the point $y$.
No independent loop variable is obtained by considering
a gauge invariant non local variable containing $E(x)$
as an insertion. 

It is important to recall, once again,  that in this
approach all the information about the theory is encoded in terms of
loop variables,  which are  gauge invariant under small and large gauge
transformations. That means, in particular, that the electromagnetic
information is encoded in the elements of the U(1) group $U(\eta_x^y)$
and consequently the loop representation naturally describes compact
electrodynamics.  In other words, since the basic electromagneric variable
is ${\rm exp} \ ie \int_0^L dx A(x) $, it is enough to restrict $ \int_0^L dx A(x)$ to the interval
$[0, \frac{2 \pi}{e}]$.

The induced non zero Poisson brackets
among the loop variables are
\begin{eqnarray}
\left\{ T^0(\gamma), E(x) \right\} &=&
      ie \oint_{\gamma}dz\,\delta(x,z)\, T^0(\gamma), \label{eq:te}\\
\left\{ \Pi_i(\eta_x{}^y), E(z) \right\} &=&
       ie \int_{\eta_x{}^y} du\, \delta(z,u)\,
       \Pi_i(\eta_x{}^y)\,,\;\;\;i=0,1,2,3, \\
\left\{ \Pi_0(\alpha_x{}^y), \Pi_1(\eta_u{}^v) \right\} &=&
      { i\over \hbar}
\delta(x,v)\, \Pi_0 \left( (\eta\circ\alpha)_u{}^y \right), \\
\left\{ \Pi_0(\alpha_x{}^y), \Pi_2(\eta_u{}^v) \right\} &=&
       -{i\over \hbar}
 \delta(u,y)\, \Pi_0 \left( (\alpha\circ\eta)_x{}^v \right), \\
\left\{ \Pi_0(\alpha_x{}^y), \Pi_3(\eta_u{}^v) \right\} &=&
   {i\over \hbar} \delta(x,v)\, \Pi_2 \left(
(\eta\circ\alpha)_u{}^y \right),
\nonumber \\
&& -{i\over \hbar}
 \delta(y,u)\, \Pi_1 \left((\alpha\circ\eta)_x{}^v \right),  \\
\left\{ \Pi_i(\alpha_x{}^y), \Pi_i(\eta_u{}^v) \right\} &=&
      - {i\over \hbar}
 \delta(y,u)\, \Pi_i \left( (\alpha\circ\eta)_x{}^v \right) +
\nonumber\\
&&{i\over \hbar}
 \delta(x,v)\, \Pi_i \left((\eta\circ\alpha)_u{}^y \right),
     i=1,2, \\
\left\{ \Pi_1(\alpha_x{}^y), \Pi_3(\eta_u{}^v) \right\} &=&
   {i\over \hbar}  \delta(x,v)\, \Pi_3 \left(
(\eta\circ\alpha)_u{}^y \right),
\\
\left\{ \Pi_2(\alpha_x{}^y), \Pi_3(\eta_u{}^v) \right\} &=&
    - {i\over \hbar}
\delta(y,u)\, \Pi_3 \left( (\alpha\circ\eta)_x{}^v \right).
      \label{eq:p2p3}
\end{eqnarray}
Since it will be crucial later on, we stress here that the line integral
$\int_{\eta_x{}^y} du\, \delta(u,z) $ is more conveniently expressed
as \cite{HallinLiljenberg}
\begin{equation}
\theta(x,y;z)\equiv \int_{\eta_x{}^y} du\, \delta(u,z)
= \frac{1}{L} \left( y-x + \sum_{k\ne 0} \frac{1}{ik}
                   \left( e^{ik(y-z)} - e^{ik(x-z)} \right)
                   \right)\, ,
\label{THETA}
\end{equation}
where $k=\frac{2\pi n}{L} , \ n=\pm 1 , \pm 2 , \dots$.
We observe that
$\lim_{x\rightarrow y} \theta(x,y;z) = 0$. We can  verify also  that
\begin{equation}
\frac{\partial\theta(x,y;z)}{\partial z} =
\delta(x,z) - \delta(y,z) \;\; .
\end{equation}
The representation of the delta function is given by
\begin{equation}
\delta(x,y)={1\over L} \sum_{n=-\infty}^{+ \infty}
e^{{2 \pi i n \over L}(x-y)}
\label{DELTA}
\end{equation}
and the following properties can be directly obtained
\begin{equation}
\delta(x,y)=\delta(y,x), \quad \delta(x,0)=\delta(x,L).
\label{PROPDEL}
\end{equation}

To proceed further we rewrite   ${\cal H, G}, \rho$ in terms
of loop variables
\begin{eqnarray}
{\cal H}(x) &=& \frac{1}{2} E^2(x)
+ \frac{i\hbar}{2} \lim_{y\rightarrow x}
(\partial_x - \partial_y)(\Pi_2 (\alpha_y{}^x) - \Pi_1 (\alpha_y{}^x))
\label{eq:dham}  \\
{\cal G}(x) &=& \partial_x E(x) - e \hbar \lim_{y\rightarrow x}
     \left( \Pi_1 (\alpha_y{}^x) + \Pi_2 (\alpha_y{}^x) \right)
     \label{eq:gauss} \\
\rho(x) &=& \lim_{y\rightarrow x}\,e\,
     \left( \Pi_1 (\alpha_y{}^x) + \Pi_2 (\alpha_y{}^x) \right)
     \label{eq:dnum}
\end{eqnarray}
Although the quantities ${\cal H}, \rho$ are gauge invariant by
construction
in terms of the local fields,
one can directly verify that their Poisson brackets with the
Gauss constraint are identically zero using the above algebra, i.e.
Eqs. (\ref{eq:te}-\ref{eq:p2p3}).
Because it is illuminating we next show that the charge
$Q=\int_0^L dw \rho(w)$ is a constant of motion
\begin{eqnarray}
\left\{ Q, H\right\} &=& e  \int_0^L dw \int_0^L dy
  \lim_{\stackrel{\scriptstyle z\rightarrow w}
  {x \rightarrow y}}
  \left[ E(y) ie \theta(z,w;y) ( \Pi_1(\eta_z{}^w)
                                +\Pi_2(\eta_z{}^w) ) \right. \nonumber\\
 & &  + i\partial_y
     ( -i \delta(w,x)\, \Pi_2((\eta\circ\alpha)_z{}^y)
       +i \delta(z,y)\, \Pi_2((\alpha\circ\eta)_x{}^w) )  \nonumber \\
 & &  \left. - i\partial_y
     ( -i \delta(w,x)\, \Pi_1((\eta\circ\alpha)_z{}^y)
       +i \delta(z,y)\, \Pi_1((\alpha\circ\eta)_x{}^w) )\right] =0
  \,.
\label{eq:nh}
\end{eqnarray}
The Hamiltonian  is $H =\int_0^L dx  \ {\cal H}$. Note that the
first line of
(\ref{eq:nh}) is zero because of
$\lim_{z\rightarrow w} \theta(z,w;y)=0$, provided
$\lim_{z\rightarrow w} \Pi_+(z,w)$ is finite.
That the second  line is also zero can be  readily seen as follows:
after taking the limits  $x\rightarrow y\,,\,z\rightarrow w$
in the term $\partial_y\delta(z,y)\, \Pi_2((\alpha\circ\eta)_x{}^w)$ and
integrating  by parts
the first term
$\delta(w,y)\, \partial_y\Pi_2(\alpha_w{}^y)$ gets cancelled.
The remaining boundary term is zero after performing the second integral.
The third line is zero for analogous reasons.


\section{Quantum framework}
Our general procedure of quantization will be to promote the
observables $A,B$  to operators $\hat A,\hat B$ and
let their  Poisson brackets $\left\{A,B\right\}=C$
go over (anti)commutators
$[\hat{A},\hat{B}]_{\pm} = i\hbar
\hat C $ according to the standard prescription.

\subsection{Local field representation}
We define this representation by choosing  the configuration
variables as  \
$\hat{A}(x), \  \hat{\psi}_1^*(x), \\  \hat{\psi}_2(x)$
acting multiplicatively on a wave function
$\Psi(A(x), \psi_1^*(x), \psi_2(x))$.
The remaining variables will be represented by:
\begin{equation}
\hat{E}(x)=-i\hbar \frac{\delta}{\delta A(x)},\;\;\;
\hat{\psi}_1= \frac{\delta}{\delta \psi_1^*(x)},\;\;\;
\hat{\psi}_2^*= \frac{\delta}{\delta \psi_2(x)}\;.
\end{equation}
In this representation,  the state of nothing $|0 \rangle$  which
does not contain any field excitation,
is given by a constant functional.

\subsection{Loop representation}

This representation will be constructed as a quantum  realization of
the loop variables Poisson algebra given by
Eqs.(\ref{eq:te}-\ref{eq:p2p3}).
To this end, it is convenient to first characterize the  state of nothing
$|0 \rangle$, which is not to be confused with the physical vacuum of
the theory.  The former is characterized by  $\langle A, \psi^*_1,
\psi_2|0
\rangle= {\rm  const.}$ in the previous representation. This leads
us to define the state of nothing  as
\begin{equation}
\hat{E}(x)|0\rangle=0,
\quad \hat{\Pi}_i(x,y)|0\rangle=0, \quad i=1,2,3.
\end{equation}
Note that these conditions  enforce the gauge
invariance of this state. The above definitions imply that, in the
connection representation, the functional derivative operator must always
go to the right. In this way,  we  choose  the following
representation for the
gauge invariant operators $ {\hat \Pi}_i(\eta_x{}^y)$
\begin{eqnarray}
{\hat \Pi}_0(\eta_x{}^y) &=&\psi_1^*(x) U(\eta_x{}^y)
\psi_2(y), \\
{\hat \Pi}_1(\eta_x{}^y) &=& \psi_1^*(x) U(\eta_x{}^y)
{\delta\over \delta \psi_1^*(y)}, \\
{\hat \Pi}_2(\eta_x{}^y) &=&-\psi_2(y) U(\eta_x{}^y){\delta \over
\delta \psi_2(x)}, \\
{\hat \Pi}_3(\eta_x{}^y) &=&{\delta \over \delta \psi_2(x)}
U(\eta_x{}^y)
{\delta\over \delta \psi_1^*(y)},
\end{eqnarray}
where the minus sign in ${\hat \Pi}_2$ has been introduced to recover the
classical
limit of Eq.(\ref{PI2}). Note that
${\hat \Pi}_3^{\dagger}(x,y) = {\hat \Pi}_0(y,x)$ in the standard scalar
product of Grassmann variables..
The resulting non-zero commutators
are
\begin{eqnarray}
{[{{\hat \Pi}_{0}} (u,v),{{\hat \Pi}_{1}} (x,y)]} &=&
 - \delta (u,y) {{\hat \Pi}_{0}} (x,v),
\label{eq:p0p1}\\
{[{{\hat \Pi}_{0}} (u,v),{{\hat \Pi}_{2}} (x,y)]} &=&
\delta (x,v){{\hat \Pi}_{0}} (u,y),
\label{eq:p0p2}\\
{[{{\hat \Pi}_{0}} (u,v),{{\hat \Pi}_{3}} (x,y)]} &=&
      \delta (x,v){\hat \Pi}_{1} (u,y)
 - \delta (u,y) {\hat \Pi}_{2} (x,v)\nonumber \\
& &  -\delta(u,y)\delta(x,v),
     \label{eq:p0p3}\\
{[{\hat \Pi}_{i} (u,v),{\hat \Pi}_{i} (x,y)]} &=&
     \delta (x,v) {\hat \Pi}_{i} (u,y) -
\delta (u,y){\hat  \Pi}_{i} (x,v), \quad i=1,2
     \label{eq:ppmppm}\\
{[{\hat \Pi}_+ (u,v),{\hat \Pi}_+ (x,y)]} &=&
     \delta (x,v) {\hat \Pi}_+ (u,y) - \delta (u,y) {\hat \Pi}_+ (x,v),
     \label{eq:p+p+}\\
{[{\hat \Pi}_- (u,v),{\hat \Pi}_- (x,y)]} &=&
     \delta (x,v) {\hat \Pi}_+ (u,y) - \delta (u,y) {\hat \Pi}_+ (x,v),
     \label{eq:p-p-}\\
{[{\hat \Pi}_+ (u,v),{\hat \Pi}_- (x,y)]} &=&
     \delta (x,v) {\hat \Pi}_- (u,y) - \delta (u,y) {\hat \Pi}_- (x,v),
     \label{eq:p+p-}\\
{[{\hat \Pi}_3(u,v),{\hat \Pi}_1(x,y)]} &=&
     \delta(x,v){\hat \Pi}_3(u,y),
     \label{eq:p3p+}\\
{[{\hat \Pi}_3(u,v),{\hat \Pi}_2(x,y)]} &=&
     - \delta(u,y) {\hat \Pi}_3(x,v),
\label{eq:p3p-}\\
{[{\hat \Pi}_i(u,v),{\hat E}(x)]} &=&
     - e \hbar  \theta(u,v ; x) {\hat \Pi}_i(u,v), \ \ i =0,1,2,3.
\label{eq:pie}
\end{eqnarray}
where the c-number contribution in Eq.(\ref{eq:p0p3}) arises from  the
ordering of the operators. In the above equations we have
introduced the notation
\begin{equation}
{\hat \Pi}_{\pm}(x,y):={\hat  \Pi}_1(x,y) \pm
{\hat \Pi}_2(x,y)
\label{PIMM}
\end{equation}
and from now on  we denote
${\hat \Pi}_i(\eta_x^y)$ simply by ${\hat \Pi}_i(x,y) $.

An heuristic application of the loop transform
shows that the operators ${\hat T}^0(\gamma)$
together with ${\hat \Pi}_0(\eta_x{}^y)$ acting
on the state of nothing $|0>$, create states with closed and open curves
respectively \cite{GambiniTrias,HugoRovelli}. This
can be formally stated as
\begin{eqnarray}
{\hat T}^0(\gamma_1)\dots{\hat T}^0(\gamma_q) |0\rangle&=&
|\gamma_1,\dots, \gamma_q \rangle, \label{eq:t0t0} \\
{\hat \Pi}_0(x_1,y_1)\dots{\hat \Pi}_0(x_m, y_m)
|0\rangle&=&|x_1, y_1,\dots, x_m, y_m \rangle.
\label{PARES}
\end{eqnarray}
Hence, the commutators of the operators with $\hat \Pi_0$
will provide the action of such operators upon generic states.
For simplicity we will consider
states like  $|\gamma\rangle $ and
$|\dots, x, y, \dots \rangle$ separately.
The basic idea
is to apply the corresponding commutator to the state of nothing.
Suppose we
want to calculate $\hat{E}(x)|\gamma \rangle$. To this end let us
consider
\begin{eqnarray}
[\hat{E}(x), \hat{T}^0(\gamma)]|0\rangle
&=& e\hbar \oint_{\gamma}du\,\delta(x,u)
\hat{T}^0(\gamma)|0\rangle, \nonumber \\
\hat{E}(x)|\gamma\rangle &=& e\hbar \oint_{\gamma}du\delta(x,u)\,
|\gamma\rangle=e\hbar n|\gamma\rangle,
\end{eqnarray}
where $n$ is  the winding number of the closed curve $\gamma$.
In a completely analogous way we obtain
\begin{equation}
\hat{E}(z)|x_1, y_1,\dots,x_m, y_m\rangle
= e\hbar \left( \sum_{k=1}^m \theta(x_k,y_k;z)\right)
|x_1, y_1,\dots,x_m, y_m\rangle.
\end{equation}

Let us observe that in spite of Eq.(\ref{eq:t0t0}), which would demand
$q$ labels $n_q$, we need only  one label counting the total
winding number $n=n_1+\dots+n_q$. This is because
$\hat E(x) | n_1, n_2\rangle= e\hbar (n_1+n_2)| n_1, n_2\rangle
\approx \hat E(x) | n_1 + n_2\rangle$, which allow us  to identify the
states
$| n_1, n_2\rangle $ and $| n_1 + n_2\rangle $ up to a phase. Thus,
we identify
$|\gamma_1,\dots, \gamma_q \rangle =| n \rangle$.

The remaining operators $\hat{\Pi}_i, \ i=1,2,3$, leave invariant
the state associated to the closed curve $\gamma$, while some examples of
their
action upon  states defined by open curves are
\begin{eqnarray}
\hat{\Pi}_1(w, u)| x , y\rangle &=&
\delta(x,u) | w, y \rangle, \nonumber  \\
\hat{\Pi}_2(w, u)|x, y\rangle &=&
- \delta(y,w) | x, u \rangle, \nonumber \\
{\hat \Pi}_3(u,v)|x,y\rangle &=& \delta(x,v) \delta(u,y) |0\rangle \,,
\nonumber \\
\hat{\Pi}_3( u, v)| x, y,  w, z\rangle &=&
- \delta(x,v)\delta(z,u)
| w, y \rangle - \delta(y,u) \delta(w,v)
| x, z \rangle\nonumber\\
&& + \delta(u,y)\delta(x,v)| w, z \rangle  +\delta(u,z)\delta(v,w)
| x, y \rangle
\end{eqnarray}

In this way, we consider our Hilbert space to be spanned by the set
of all
vectors
\begin{equation}\label{BDEF}
 |n; x_1y_1, \dots, x_a , y_a \rangle \equiv |n\rangle \otimes
|x_1y_1, \dots, x_a , y_a \rangle, \   -\infty< n< +\infty,  \
a=0,1,2 \dots,
\infty.
\end{equation}
These vectors satisfy the following orthogonality and closure
properties
\begin{eqnarray}\label{ORTH}
\langle n;  x_1, y_1, \dots,x_a, y_a && |m ; u_1v_1, \dots, u_b ,
v_b \rangle =
\delta_{m,n} \delta_{a,b} \nonumber \\
&& \sum_{q_i,p_j} \epsilon_{q_1q_2 \dots q_a}\epsilon_{p_1p_2 \dots p_a}
\delta( x_{1} - u_{q_1}) \delta( y_{1} - v_{p_1})
\dots
\delta( x_{a} - u_{q_a}) \delta( y_{a} - v_{p_a}),
\end{eqnarray}
\begin{eqnarray}\label{CLOS}
\sum_{n=-\infty}^{+\infty}\ \sum_{a=0}^{+\infty} \int dx_1dy_1
\dots dx_ady_a \
{1\over {a !}^2}
| n;  x_1, y_1, \dots, x_a, y_a \rangle \langle  n;  x_1, y_1,
\dots,x_a, y_a |
= 1,
\end{eqnarray}
which are a direct consequence of the basic algebra
(\ref{eq:p0p1})-(\ref{eq:p3p-}).
They also satisfy the following exchange properties
\begin{eqnarray}
|n;\dots,x_i,y_i,\dots,x_j,y_j, \dots x_ay_a\rangle &=&
    |n;\dots,x_j,y_j,\dots,x_i,y_i, \dots, x_a,y_a\rangle,\label{IJIJ} \\
 |n;\dots,x_i,y_i,\dots,x_j,y_j, \dots x_ay_a\rangle&=&
-|n;\dots,x_j,y_i,\dots,x_i,y_j, \dots x_ay_a\rangle,\label{XIJ}    \\
|n;\dots,x_i,y_i,\dots,x_j,y_j, \dots x_ay_a\rangle &=&
-|n;\dots,x_i,y_j,\dots,x_j,y_i, \dots x_ay_a\rangle.\label{YIJ}
\end{eqnarray}

Equation (\ref{IJIJ})  is a direct consequence of the definition in Eq.
(\ref{PARES})  and  the fact that
$[\hat{\Pi}_0(x_i,y_i),\hat{\Pi}_0(x_j,y_j)]=0$ , while  Eqs.
(\ref{XIJ}) and (\ref{YIJ}) arise from the anticommuting property of the
fermion operators together with the abelian composition rule
\begin{equation}\label{PRODU}
U(u,v)U(x,y)=U(u,y)U(x,v).
\end{equation}

In the loop representation,  a general  wave
function  $|\Psi\rangle$  can be written as
\begin{equation}
|\Psi\rangle= \sum_{a=0}^{\infty}  \sum_{n=-\infty}^{+ \infty}\int
dx_1dy_1
\dots dx_ady_a {1\over a !} \Psi^a(n; x_1, y_1, \dots, x_a, y_a )
|n; x_1y_1,
\dots, x_a , y_a \rangle, \label{eq:wf}
\end{equation}
where
\begin{equation}
\langle n;  x_1, y_1, \dots,x_a, y_a| \Psi \rangle =  \Psi^a(n; x_1, y_1,
\dots, x_a, y_a )
\label{COMP}
\end{equation}
are the corresponding components. They inherit the exchange properties
(\ref{IJIJ}), (\ref{XIJ}) and (\ref{YIJ}) of the basis vectors and also
satisfy  the following boundary conditions
\begin{eqnarray}
\Psi^a(n;\dots,x_k,y_k+mL,\dots)   &=&
    e^{im\pi} \Psi^a (n+m;\dots,x_k,y_k,\dots),\label{eq:yp}    \\
\Psi^a(n;\dots,x_k+sL,y_k,\dots)   &=&
    e^{-is\pi} \Psi^a (n-s;\dots,x_k,y_k,\dots),\label{eq:xp}
\end{eqnarray}
which  arise from the property
$\Pi_0(x,y+mL) = e^{im\pi} \psi_1^*(x) U(x,y)T^0 (m)
\psi_2(y)$, in virtue of  the  boundary conditions
(\ref{BC}). As a consequence of the symmetry properties
(\ref{eq:yp}) and (\ref{eq:xp})  we will assume from here on that
$0 \leq x_k, y_k < L$.

The scalar product  is given by
\begin{eqnarray}\label{PRODI}
\langle \Phi | \Psi \rangle = \sum_{a=0}^{\infty}
\sum_{n=-\infty}^{+ \infty}
\sum_{i_a}\int dx_1dy_1 \dots dx_ady_a &&(\Phi^a(n; x_1, y_1,
\dots, x_a, y_a
))^*\nonumber  \\
&&\times  \Psi^a(n; x_1, y_{i_1}, \dots, x_a, y_{i_a} ),
\end{eqnarray}
where  the functions $ \Phi^a(n; x_1, y_1, \dots, x_a, y_a )$ are the
components of  the wave function $ | \Phi\rangle$.

In order to find the associated  Schroedinger equation we need to
compute
the action of the operators appearing in the Hamiltonian density
(\ref{eq:dham}) upon  the basis vectors. We
obtain
\begin{eqnarray}
\hat{E}^2(x) |n;x_1,y_1,\dots,x_a,y_a \rangle &=&
    e^2 {\hbar^2} \left[ n+\sum_{k=1}^a \theta(x_k,y_k,x)\right]^2
    |n;x_1,y_1,\dots,x_a,y_a\rangle, \label{eq:e2} \\
\hat{\Pi}_1(x,y) |n;x_1,y_1,\dots,x_a,y_a\rangle  &=&
    \hbar\sum_{k=1}^a \delta(x_k,y)
    |n;\dots,x_{k-1},y_{k-1},x,y_k,x_{k+1},y_{k+1},\dots,
x_a,y_a\rangle, \nonumber\\
    & & \label{eq:pi1xy} \\
\hat{\Pi}_2(x,y) |n;x_1,y_1,\dots,x_a,y_a\rangle  &=&
    -\hbar\sum_{k=1}^a \delta(y_k,x)
    |n;\dots,x_{k-1},y_{k-1},x_k,y,x_{k+1},y_{k+1},\dots, x_a,
y_a\rangle. \nonumber\\
   & &  \label{eq:pi2xy}
\end{eqnarray}
Using the above actions, we have explicitly verified that the basis
vectors
(\ref{BDEF}) are
annihilated by the Gauss law constraint (\ref{eq:gauss}).

The Hamiltonian  $\hat{H}$ is block-diagonal  in the subspace of
fixed number
of pairs and fixed ${n}$. Thus we look for solutions of the Schroedinger
equation,  ${\hat  H} |\Psi\rangle = E |\Psi\rangle$,  which are of the
form
\begin{equation}\label{WFSS}
|\Psi\rangle_{a,n}= \int dx_1dy_1 \dots dx_ady_a  \Psi^a(n; x_1,
y_1, \dots,
x_a, y_a ) |n; x_1y_1, \dots, x_a , y_a \rangle,
\end{equation}
The action upon the components of the wave function is
\begin{eqnarray}
\langle n;x_1,y_1, &&\dots,x_a,y_a|{\hat H}|\Psi\rangle_{a,n} =
   { e^2 {\hbar^2} \over 2}
 \left( \int_0^{L}dx\left[ n+\sum_{k=1}^a
 \theta(x_k,y_k,x)\right]^2 \right) \Psi^a
(n;x_1,y_1,\dots,x_a,y_a)\nonumber
\\
& & + i \frac{\hbar}{2} \int_0^{L} dx
    \sum_{k=1}^a \partial_x\delta(x_k,x)
\Psi^a(n;\dots,x_{k-1},y_{k-1},x,y_k,x_{k+1},y_{k+1},\dots, x_a,
y_a)\nonumber
\\
& & +  i \frac{\hbar}{2}\int_0^{L} dx \sum_{k=1}^a \delta(y_k,x)
    \partial_x
\Psi^a(n;\dots,x_{k-1},y_{k-1},x_k,x,x_{k+1},y_{k+1},\dots, x_a,
y_a )
    \nonumber \\
& & -i \frac{\hbar}{2} \int_0^{L} dx
    \sum_{k=1}^a \partial_x\delta(y_k,x)
    \Psi^a(n;\dots,x_{k-1},y_{k-1},x_k,x,x_{k+1},y_{k+1},\dots, x_a,
y_a)\nonumber \\
& & -i \frac{\hbar}{2}\int_0^{L} dx \sum_{k=1}^a \delta(x_k,x)
    \partial_x
\Psi^a(n;\dots,x_{k-1},y_{k-1},x,y_k,x_{k+1},y_{k+1},\dots, x_a,
y_a).
\nonumber\\
& & \label{eq:ham1}
\end{eqnarray}
Integrating by parts the term containing the derivative of the
delta function in (\ref{eq:ham1})
we obtain the final result
\begin{eqnarray}
\langle n;x_1,y_1, \dots,x_a,y_a|{\hat H}|\Psi\rangle_{a,n} &=&
   { e^2 {\hbar^2} \over 2}
 \left( \int_0^{L}dx\left[ n+\sum_{k=1}^a
 \theta(x_k,y_k,x)\right]^2 \right)
    \Psi^a(n;x_1,y_1,\dots,x_a,y_a)\nonumber  \\
& & -i\hbar \sum_{k=1}^a \left(\frac{\partial}{\partial x_k}
     - \frac{\partial}{\partial y_k} \right)
 \Psi^a(n;x_1,y_1,\dots,x_k,y_k,\dots, x_a,y_a).\nonumber \\
\label{eq:ham2}
\end{eqnarray}

\section {External field analysis}

As a first  step in the quantization of the full system we consider  the
quantization of the fermionic
fields  in a background
electromagnetic field. According to Ref. \cite{IsoMurayama}, the
fermionic
field operators are given by
\begin{equation}
\psi_1(x,t)=\sum_{n} a_n \phi_n(x)e^{-{i\over \hbar}\epsilon_n t}, \quad
\psi_2(x,t)=\sum_{n} b_n^{\dagger} \phi_n(x)
e^{{i\over \hbar}\epsilon_n t},\label{psi12}
\end{equation}
where we have slightly changed the notation in the second equation
(\ref{psi12}).
The operators $a_n, b_n$ are standard fermionic  annihilation operators
satisfying the non-zero anticommutators: \ $\{a_n, a_m^\dagger\}=
\delta_{mn}=\{b_m, b_n^\dagger\}$.
The basic wave functions $\phi_n$ , together with the
eigenvalues of the energy are given by
\begin{equation}
\phi_n(x)={1\over\sqrt L}e^{{i\over \hbar}\epsilon_nx -ie\int_0^x
A(z)dz},\quad
{1\over \hbar}\epsilon_n=
{2\pi\over L}\left(n+{1\over2} +{eL\over 2\pi}c\right)\equiv {2\pi
n\over L} + \theta,
\label{VFP}
\end{equation}
where
\begin{equation}
c={1\over L}\oint A(z)dz, \quad \theta={\pi\over L}+ ec .
\label{cteta}
\end{equation}
The energy eigenvalues are invariant under small gauge transformations
in such a way that the corresponding eigenfunctions transform covariantly
according to (\ref{GGT}). The same structure is kept  in the case
of large gauge transformations. To see this,   it is enough to  
recall that
in the compact U(1) case under consideration
$c$ lies in the interval $[ 0, {2 \pi \over e L}]$. This means that  
  $c$  is
invariant under large gauge transformations.

In the scalar product
\begin{equation}
(\phi, \xi)=\int_0^L dz \  \left[\phi(z)\right]^*\xi(z),
\label{PE}
\end{equation}
the functions defined in (\ref{VFP}) are
orthonormal, i.e. \ $(\phi_m, \phi_n)=\delta_{mn}$.

It is convenient to introduce  a bi-local Fourier transform of the loop
space operators $\hat\Pi(x,y)$ in
the following way
\begin{equation}
\Pi^{mn}=\int {dx dy\over L} e^{{i\over \hbar}\epsilon_m x}
e^{-{i\over \hbar}\epsilon_n y}\hat\Pi(x,y),
\label{FT}
\end{equation}
together with its inverse
\begin{equation}
\hat\Pi(x,y)={1\over L}\sum_{m,n} e^{-{i\over \hbar}\epsilon_m x}
e^{{i\over \hbar}\epsilon_n y}\Pi^{mn}.
\label{IFT}
\end{equation}

Also,  we define the Fourier transform of the electric field
operator $E_b, \
$
as
\begin{equation}\label{EMODES}
E(x) =  \sum_{b=-\infty }^{\infty} E_b \  { e}^{ -{2\pi i  b \over L} x},  \quad E^\dagger_b
=E_{-b}, \quad   b=0,\pm1, \pm 2, \dots
\end{equation}
which leads to the following inverse transformations
\begin{equation}\label{INVEMOD}
 E_b= {1\over L} \int_0^L
dx  E(x) \ { e}^{ {2\pi i  b \over L} x} .
\end{equation}

The commutator algebra (3.7)-(3.13) can be directly rewritten in
terms of the
Fourier-transformed
operators $\Pi^{mn}$ and $E_b$. The result
is
\begin{eqnarray}
\left[ \Pi_0^{mn} , \ \Pi_1^{kl} \  \right]&=&  - \delta^{ml} \Pi_0^{kn},
\label{eq:crf1}\\
\left[ \Pi_0^{mn} , \ \Pi_2^{kl} \  \right]&=&  \  \ \delta^{kn}
\Pi_0^{ml}
, \label{eq:crf2}\\
\left[ \Pi_i^{mn} , \ \Pi_i^{kl} \  \right]&=& \  \delta^{kn}
\Pi_i^{ml}  -
\delta^{ml} \Pi_i^{kn}  , \quad  i=1,2,  \label{eq:crf3}\\
\left[ \Pi_+^{mn} , \ \Pi_-^{kl} \  \right]&=& \  \delta^{kn}
\Pi_-^{ml}  -
\delta^{ml} \Pi_-^{kn}  ,  \label{eq:crf4}\\
\left[ \Pi_3^{mn} , \ \Pi_1^{kl} \  \right]&=&   \ \   \delta^{kn}
\Pi_3^{ml},    \label{eq:crf5}\\
\left[ \Pi_3^{mn} , \ \Pi_2^{kl} \  \right]&=&  - \delta^{ml} \Pi_3^{kn},
\label{eq:crf6}\\
\left[ \Pi_0^{mn} , \ \Pi_3^{kl} \  \right]&=& \  \delta^{kn}
\Pi_1^{ml}  -
\delta^{ml} \Pi_2^{kn}
-\delta^{kn} \delta^{ml}\,,  \label{eq:crf7}
\end{eqnarray}
where $\Pi_3^{mn\,\dagger}=\Pi_0^{nm}$.

The commutators involving the electric modes are
\begin{eqnarray}\label{CEM}
 \left[  E_a, \ E_b  \right] &=&0, \quad \left[  c, \ E_b  \right] =
{i\hbar\over L} \delta_{b0},
 \nonumber \\
\left[  T^0(n), \ E_b  \right] &=&- {en\hbar} \  T^0(n)
\delta_{b0}, \quad
\left[  \Pi_i^{kl}, \ E_0  \right] =0, \quad  i= 0, 1,2,3,  \nonumber \\
\left[  \Pi_i^{kl}, \ E_b  \right]&=& {ie\hbar\over 2\pi b }
\left( \Pi_i^{k \ l-b} -  \Pi_i^{k+ b \ l}   \right),
b\neq 0.
\end{eqnarray}

\subsection{Vacuum state in a background electromagnetic field}

The Hamiltonian density  that describes the external field
approximation is
given by the
the second term in the RHS of  (\ref{eq:dham}).  When written in the
momentum
space, the corresponding Hamiltonian is
\begin{equation}\label{mopmham}
H_{D}= \sum_m \epsilon_m \Pi_-^{mm}.
\end{equation}
The calculation of the commutators
\begin{equation}
[ \Pi_0^{kl}, \  H_{D}]=-(\epsilon_k + \epsilon_l) \ \Pi_0^{kl}, \ \
[ \Pi_3^{kl}, \  H_{D}]=+(\epsilon_k + \epsilon_l)  \ \Pi_3^{kl},
\end{equation}
implies that $ \Pi_0^{kl}$ is an energy raising operator, i.e.
$\Pi_0^{kl} \ |
E\rangle \sim | E +
\epsilon_k + \epsilon_l \rangle$, for any Hamiltonian eigenstate $  | E
\rangle$. For analogous reasons,   $ \Pi_3^{kl}$ is the
corresponding lowering
operator.  The charge operator $Q$ commutes with  $ \Pi_0^{kl}$,
showing that
$ \Pi_0^{kl}$ creates  a zero charge pair of
particles having the corresponding energies  $\epsilon_k$ and
$\epsilon_l$.
Thus we can identify
each superindex  of   $ \Pi_0^{kl}$ with a definite and opposite
charge label.
In this way we have that   $ \Pi_0^{kl} \Pi_0^{kr} | E\rangle \sim  | E +
2\epsilon_k + \epsilon_l + \epsilon_r \rangle $.
Since in one spatial dimension the momentum is proportional to the
energy, we
conclude that the state $| E + 2\epsilon_k + \epsilon_l +
\epsilon_r \rangle$
contains two fermions having the same quantum numbers  and
therefore must be
zero according to the Pauli principle. Since the eigenstates of the
Hamiltonian
provide a basis for the Hilbert space, we must have the operator
identity
\begin{equation}\label{PAPRI}
 \Pi_0^{kl} \Pi_0^{kr} =0
\end{equation}
and analogously, when the repeated indices are those in the right.

The vacuum state $|0\rangle_D$  corresponds to a filled Dirac sea
with zero
charge
which can be defined  as
\begin{equation}
| 0  \rangle_D=\prod_{k=-\infty}^{N-1} \Pi_0^{kk}|0\rangle,\quad
{}_D\langle 0|= \langle 0|\prod_{k=-\infty}^{N-1}\Pi_3^{kk},
\label{VAC}
\end{equation}
This means that all energy levels below $\epsilon_N$ are completely
filled. Provided that $-(N+{1\over 2}) \leq {eL\over 2\pi}c \leq
-(N-{1\over
2})$,
we have that  $ \epsilon_{N}\geq 0 $ and  the above
construction  includes an infinite set of negative-energy states
$\epsilon_{N-1}\leq 0 $.

{}From now on we will use the convention that  all indices ranging from
$-\infty$ to $N-1$ will
be denoted by capital letters from the beginning of the alphabet
$(A, B, C,
\dots)$, while those
going from $N$ to $+\infty$ will be denoted by lower case greek
letters from
the beginning of the
alphabet   $( \alpha, \beta, \gamma, \dots)$.

The commutation relations (\ref{eq:crf1}), (\ref{eq:crf2})
imply that
\begin{eqnarray}
\Pi_1^{kk} \left(\Pi_0^{mm}\right)|0\rangle &=&
\delta^{mk} \left(\Pi_0^{mm}\right)|0\rangle \\
\Pi_2^{kk} \left(\Pi_0^{mm}\right)|0\rangle &=&
-\delta^{mk} \left(\Pi_0^{mm}\right)|0\rangle,
\end{eqnarray}
which allow us to prove that the vacuum satisfies
\begin{equation}\label{VAC1}
\Pi_1^{AA}| 0 \rangle_D= |0 \rangle_D\,,\quad
\Pi_2^{AA}|0\rangle_D = -|0\rangle_D, \quad
 \label{vac1}
\end{equation}
\begin{equation}\label{VAC2}
\Pi_1^{\alpha\alpha}|0\rangle_D=0= \Pi_2^{\alpha\alpha} |0\rangle_D.\,
\label{vac2}
\end{equation}

Next we discuss some additional  properties of the vacuum. To begin
with let us
 calculate
$\Pi_+^{AB}|0\rangle_D$.
We have already shown that
$\Pi_+^{AA}|0\rangle_D = 0\,, \forall A$.
Now, for the case $A\ne B$, we have
\begin{eqnarray}
\Pi_+^{AB}|0\rangle_D &=& \Pi_+^{AB} \prod_{C=-\infty}^{N-1}
\Pi_0^{CC}|0\rangle
= \left[\Pi_+^{AB}\,, \prod_{C=-\infty}^{N-1} \Pi_0^{CC} \right]
|0\rangle
\nonumber \\
&=& \sum_{D=-\infty}^{N-1} \prod_{s=-\infty}^{D-1} \Pi_0^{CC}
\left[\Pi_+^{AB}\,, \Pi_0^{DD}\right] \prod_{G=D+1}^{N-1}
\Pi_0^{GG}|0\rangle.
=0
\end{eqnarray}
The last equality holds by virtue of
the commutation relation
\begin{equation}
[ \Pi_+^{AB}, \Pi_0^{CD}]= \delta^{BC}\Pi_0^{AD}-\delta^{AD}\Pi_0^{BC},
\end{equation}
together with the fact that for $A\ne B$ one will always find
products of the type $\Pi_0^{AA}\Pi_0^{AB}$, $\Pi_0^{AA}\Pi_0^{BA}$,
which are identically zero, according to the property
(\ref{PAPRI}). Thus, we
have proved
that $\Pi_+^{AB}|0\rangle_D =0$. Analogously one can show that
$\Pi_+^{\alpha\beta}|0\rangle_D=0 $. Nevertheless, both states
$\Pi_+^{A\beta}|0\rangle_D $
together with $\Pi_+^{\alpha B}|0\rangle_D $ are different from
zero and are in
fact orthogonal
to $|0\rangle_D$. Thus, we have
\begin{equation}
{}_D\langle 0|\Pi_+(x,y)|0\rangle_D =0.\label{pi+vac}
\end{equation}

Next, we repeat the calculation for  $\Pi_-(x,y)$. Again, we start
from the "momentum-space" formulation. The properties (\ref{VAC1}) ,
(\ref{VAC2})
imply
\begin{equation}
\Pi_-^{AA}| 0 \rangle_D= 2|0\rangle_D, \quad
\qquad
\Pi_-^{\alpha\alpha}|0\rangle_D=0,
\end{equation}
for the diagonal terms.When $ A\neq B$, following
analogous steps to the previous case, we find
\begin{equation}
\Pi_-^{AB}|0\rangle_D = \sum_{C=-\infty}^{N-1} \prod_{D=-\infty}^{C-1}
\Pi_0^{DD} \left[\Pi_-^{AB}\,, \Pi_0^{CC}\right] \prod_{G=C+1}^{N-1}
\Pi_0^{GG}|0\rangle.\label{pi-vacmom}
\end{equation}
The corresponding commutator here is
\begin{equation}
[ \Pi_-^{AB}, \Pi_0^{CD}]= \delta^{BC}\Pi_0^{AD}+\delta^{AD}\Pi_0^{BC}.
\end{equation}
Once more, we obtain $\Pi_-^{AB}|0\rangle_D=0, A \neq B $ because of the
presence of products of $\Pi_0 's$ having a repeated index. In
analogous way
one obtains the remaining actions leading to
\begin{equation}
\Pi_-^{AB}| 0 \rangle_D= 2 \delta^{AB}|0\rangle_D, \quad
\Pi_-^{\alpha\beta}| 0 \rangle_D=0, \quad \Pi_-^{A\beta}| 0
\rangle_D\neq 0,
\quad \Pi_-^{\alpha B}| 0 \rangle_D \neq 0.
\label{PI-VAC}
\end{equation}
Again, the non-zero vectors resulting from the last two actions in
the above
equation are orthogonal to the Dirac vacuum. In this way,
going back to the coordinate representation we obtain
\begin{equation}
{}_D\langle0|\Pi_-(x,y)| 0 \rangle_D=2 e^{-i\theta (x-y)}
{1 \over L} \sum_{A=-\infty}^{N-1} e^{-{2\pi iA\over
L}(x-y)}{}_D\langle 0
| 0 \rangle_D ={2\over L} e^{-i\theta (x-y)} F(x,y){}_D\langle 0| 0
\rangle_D.
\label{pi-vac}
\end{equation}
 In the above equation we have introduced the
function $F$ defined  as
\begin{equation}
F(x,y)=\sum_{A =-\infty}^{N-1} e^{-{2\pi iA\over L}(x-y)}=
{e^{-{2\pi i (N-1) \over L}(x-y)}
\over (1-e^{{ 2\pi i\over L}(x-y)})}, \label{FN}
\end{equation}
where the summation can be calculated because it is a geometric series.

\subsection { The vacuum energy}

Let us recall  that the Hamiltonian is given by
\begin{equation}
H_D= {i\hbar \over 2}\int _0^Ldx \lim_{y\rightarrow x}
\left[ (\partial_y - \partial_x)\Pi_-(y,x) \right].
\label{HR1}
\end{equation}
According to the relations (\ref{PI-VAC}) and  (\ref{FN}) the action of
$\Pi_-(y,x)$ on the Dirac vacuum can be written as
\begin{equation}
\Pi_-(x,y)| 0 \rangle_D={2\over L} e^{-i\theta (x-y)} F(x,y)| 0
\rangle_D+ | x,
y; -\rangle,
\end{equation}
where the state $ | x, y; -\rangle$ does not contributes in the limit of
Eq.(\ref{HR1}).
The corresponding  vacuum energy in the external field  is given by
\begin{equation}
E_D(\epsilon)={\langle 0|\hat H_D|0\rangle_D \over
\langle 0| 0\rangle_D}.
\end{equation}
The
function $E_D(\epsilon)$
will have an expansion in powers  of $\epsilon$ of the form
$E_D(\epsilon)= {a\over \epsilon^2}+b\epsilon^0 + O(\epsilon)$. We
will take
$b$ as the regularized expression for the vacuum energy $E_D$.
The resulting  term is
\begin{equation}
E_D(\epsilon) =2i\hbar \partial_{\epsilon}\left[
e^{-i\theta\epsilon}{e^{-{2 \pi i(N-1) \over L }\epsilon}  \over
1- e^{{2 \pi i \over L }\epsilon}} \right].
\end{equation}
The finite part of the above equation, when $\epsilon \rightarrow 0$,
is
\begin{equation}\label{VACEN}
E_D=\hbar \left[ {2 \pi N^2 \over L }-{2 \pi N \over L }+ 2N \theta -
\theta + {\pi  \over 3L } + {\theta^2 L \over 2 \pi}\right],
\end{equation}
which coincides with the result of Ref.\cite{IsoMurayama}. In the  
sequel we
choose $N=0$,  in such a way that  $ -\pi \leq ecL
\leq  +
\pi$,  which reinforces the fact that $ ec$ is  a compact degree of  
freedom.

\section{ The Axial Anomaly}

The vector and axial currents are defined as:
\begin{eqnarray}
J_{V}{}^{\mu}(x)&:=& \lim_{y\rightarrow x}
e\overline{\psi}(y)\gamma^{\mu} U(y,x) \psi(x), \\
J_A{}^{\mu}(x)&:=& \lim_{y\rightarrow x}
e\overline{\psi}(y)\gamma^{\mu}\gamma_5 U(y,x) \psi(x).
\end{eqnarray}
Thus, in terms of loop variables, their components become
\begin{eqnarray}
J_V{}^0 &=& \lim_{y\rightarrow x} e \Pi_+(y,x)
\;,\;\;\;\;
J_V{}^1 = \lim_{y\rightarrow x} e \Pi_-(y,x),
 \\
J_A{}^0 &=& \lim_{y\rightarrow x} e \Pi_-(y,x)
\;,\;\;\;\;
J_A{}^1 = \lim_{y\rightarrow x} e \Pi_+(y,x).
\end{eqnarray}
Next we calculate the relevant commutators in order to
determine $[\hat Q_{A,V},\hat H]$, where
\begin{equation}
\hat Q_V:=\int_0^L dx \lim_{y\rightarrow x}
e \hat\Pi_+(y,x)\;\;,\;\;\;\;\;\;
\hat Q_A:=\int_0^L dx \lim_{y\rightarrow x}
e \hat\Pi_-(y,x)\,
\end{equation}
and $\hat H$ is the full Hamiltonian.
To do so we start by looking at the following commutators
\begin{eqnarray}
[\hat\Pi_{\pm}(y,x), \hat H] =
-\frac{e\hbar}{2} \int_x^y dw (\hat
E(w)\hat\Pi_{\pm}(y,x)+\hat\Pi_{\pm}(y,x)
\hat E(w))
-i\hbar (\partial_x+\partial_y)\hat\Pi_{\mp}(y,x),  \label{eq:cp1h}
\end{eqnarray}
In obtaining the above  commutators, we have used the property
\begin{equation}
\delta(x,L)\hat\Pi_{\pm}(y,L) -\delta(x,0)\hat\Pi_{\pm}(y,0) =0=
\delta(y,L)\hat\Pi_{\pm}(L,x) -\delta(y,0)\hat\Pi_{\pm}(0,x),
\end{equation}
according to Eq. (\ref{DELTA}).
Let us consider  now $\int_0^L dx\lim_{y\rightarrow x}
[\hat\Pi_1(y,x),\hat H]$.
One can readily see that the integral
\begin{equation}
\int_0^L dx
\lim_{y\rightarrow x}(\partial_x + \partial_y) \hat\Pi_{\pm}(y,x)
= \lim_{\epsilon\rightarrow 0} \int_0^L dx
\partial_x\hat\Pi_{\pm}(x+\epsilon,x)
= \hat\Pi_{\pm}(L,L) - \hat\Pi_{\pm}(0,0) =0\,.
\end{equation}
The only term left is
\begin{equation}
\int_0^Ldx \lim_{y\rightarrow x} [\hat\Pi_{\pm}(y,x),\hat H]
= -\frac{e\hbar}{2}\lim_{\epsilon\rightarrow 0}
\int_0^L dx \int_x^{x+\epsilon} dw
[\hat
E(w)\hat\Pi_{\pm}(x+\epsilon,x)+\hat\Pi_{\pm}(x+\epsilon,x)\hat
E(w)]\, .
\label{eq:lcp1h}
\end{equation}
Thus, the above limit would yield zero provided $\hat \Pi_{\pm}(x,x)$
is finite. Nevertheless we expect $\hat\Pi_{\pm}(x,x)$ to be divergent
so that  the limit must be carefully calculated.

In order to obtain the divergence of the vector current we
start from
\begin{equation}
\partial_0 \hat J_V{}^0 = {i \over \hbar}[\hat H,\hat J_V{}^0] \,.
\end{equation}
The calculation of the commutator leads to
\begin{equation}
\partial_0 \hat J_V{}^0 =
\lim_{y\rightarrow x}{ie^2\over 2}\int_x^y dw
(\hat E(w)\hat\Pi_+(y,x)+\hat\Pi_+ (y,x)\hat E(w))-
\lim_{y\rightarrow x} (\partial_x+\partial_y)
e\hat\Pi_-(y,x).
\label{eq:d0j0}
\end{equation}
After taking the limit,
the second term in the r.h.s. in (\ref{eq:d0j0}) can be
identified as $-\partial_1 \hat J_V{}^1$. This can be shown by
going to the
momentum representation: on one hand we have that
\begin{equation}
\partial_1 \hat J_V{}^1=\partial_x \lim_{y\rightarrow x} e
\hat \Pi_-(y,x)={e\over \hbar L}\sum_{m,n}i(\epsilon_m-\epsilon_n)
e^{{i\over \hbar}(\epsilon_m-\epsilon_n)x}\Pi_-^{mn}. \label{d1j1}
\end{equation}
On the other hand, the limit appearing in Eq.(\ref{eq:d0j0})
is calculated as
\begin{equation}
\lim_{y\rightarrow x} (\partial_x+\partial_y)
e\hat\Pi_-(y,x)=\lim_{y\rightarrow x}{e\over \hbar L}\sum_{m,n}i
(\epsilon_m-\epsilon_n)
e^{{i\over \hbar}\epsilon_m x}e^{-{i\over \hbar} \epsilon_n y
}\Pi_-^{mn},
\end{equation}
which reduces exactly to Eq.(\ref{d1j1}) after taking the limit
$y\rightarrow x$.

Hence the final result in the calculation of Eq.(\ref{eq:d0j0}) is
\begin{equation}
\partial_{\mu}\hat J_V{}^{\mu}=
\lim_{y\rightarrow x}{ie^2\over 2}\int_x^y dw
(\hat E(w)\hat\Pi_+(y,x)+\hat\Pi_+(y,x)\hat E(w)) \;.
\label{vectcons}
\end{equation}
In analogous way we obtain
\begin{equation}
\partial_{\mu}\hat J_A{}^{\mu}=
\lim_{y\rightarrow x}{ie^2\over 2}\int_x^y dw
(\hat E(w)\hat\Pi_-(y,x)+\hat\Pi_-(y,x)\hat E(w)) \;.
\label{avectcons}
\end{equation}

Let us remark that Eqs. (\ref{vectcons}), (\ref{avectcons}) are full operator relations which 
describe the exact non-perturbative behavior of the divergences of the corresponding currents.

In order to recover the standard form of the axial anomaly, we assume
that only the fermions are quantized, and calculate the vacuum
expectation
values of the above divergences of the currents regarding $E(x)$ as an
external field,  together with the  vacuum  given by Eqs.(\ref{vac1},
\ref{vac2}). Using
Eq.(\ref{pi+vac}) and Eq.(\ref{pi-vac}) we obtain
\begin{equation}
{}_D\langle 0|\partial_{\mu}\hat J_V{}^{\mu} |0 \rangle_D =0,
\end{equation}
\begin{eqnarray}
\langle 0|\partial_{\mu}\hat J_A{}^{\mu} |0 \rangle_D  &=&
\lim_{y\rightarrow x}ie^2
E(x)(y-x){}_D\langle 0|\Pi_-(y,x)|0 \rangle_D,  \nonumber \\
{\langle 0|\partial_{\mu}\hat J_A{}^{\mu} |0 \rangle_D \over \langle 0|0
\rangle_D}&=&
{2\over L} ie^2 E(x) \lim_{y\rightarrow x}\left(
{(y-x) e^{{2\pi i  \over L}(y-x)}
\over (1-e^{{2\pi i\over L}(y-x)})}
\right)=-{e^2 \over \pi} E(x).
\end{eqnarray}

\section{ Anomalous commutators}

It is a general property of quantum field theory that, if a current is
conserved, the equal time commutator of its spatial and temporal
components cannot vanish \cite{SCHCOM}. However, by a naive use of the
canonical
commutation relations one finds that this commutator is equal to
zero. To obtain the correct result it is necessary to introduce a
regularization.  The non-locality of our  formalism provides a natural
regularization   and the algebra (\ref{eq:p0p1}
-\ref{eq:p3p-}) produces directly the anomalous commutators. For
example, we have
\begin{eqnarray}
\left[ J^0_V (u), J^1_V(x)\right] &=& \left[ \lim_{v\to u} e\hat
\Pi_+(v,u), \lim_{y\to x} e\hat\Pi_-(y,x) \right] \\
&=& e^2 \lim_{v\to u} \lim_{y\to x} \left( \delta(y-u) \hat \Pi_{-}(v,x)
- \delta(v-x) \hat \Pi_-(y,u) \right).
\end{eqnarray}
Taking the limits $v=u+\epsilon$ and $y=x+\epsilon$, with $\epsilon\to
0$, the vacuum expectation value of the commutator has the form
\begin{equation}
_D \langle 0| [ J_V^0 (u), J_V^1(x)] |0\rangle_D = {ie^2 \over \pi}
\delta^\prime (x-u).
\end{equation}

Also,  from the relations (\ref {eq:pie}), it  follows
that the
commutator of the currents and the
electric field is different of zero, which agree with the Gauss law
(\ref{eq:gauss}). In this case we have

\begin{equation}
\left[ J_V^0 (x), E(z) \right] = \left [ \lim_{y\to x} e \hat
\Pi_+(x,y) , \hat E (z) \right] = -e^2 \lim_{y\to x} \theta(x,y;z) \hat
\Pi_+(x,y),
\end{equation}

\begin{equation}
\left[ J_V^1 (x), E(z) \right] = \left [ \lim_{y\to x} e \hat
\Pi_-(x,y) , \hat E (z) \right] = -e^2 \lim_{y\to x} \theta(x,y;z) \hat
\Pi_-(x,y),
\end{equation}
{}From the above expressions we obtain  the vacuum expectation value
of these commutators
\begin{equation}
_D \langle 0| [ J_V^0 (x), E(z) ]|0\rangle_D = 0,
\end{equation}
\begin{equation}
_D \langle 0| [ J_V^1 (x), E(z) ] |0\rangle_D = {ie^2\hbar \over \pi}
\delta (x-z).
\end{equation}

\section{ The zero mode}

The simplest state with energy above the Dirac sea is the one associated
to the only degree of freedom of the electromagnetic field
which cannot be gauged away: the zero mode. Such a state can be
neatly described in the present formalism as follows.
Recalling that $c=\frac{1}{L}\int_{0}^{L} dz A(z)$,
and $-\frac{\pi}{eL}\leq c\leq \frac{\pi}{eL}$ it is useful to
consider the transformation
\begin{equation}
|c,x_1,y_1,\dots,x_a,y_a\rangle =
\sum_n {\rm e}^{-iec\left[nL +\sum_k(y_k-x_k)\right]}
|n, x_1,y_1,\dots,x_a,y_a\rangle\, .
\end{equation}
Accordingly, the action of the relevant operators becomes
\begin{eqnarray}
T^0 [\gamma] |c,x_1,y_1,\dots,x_a,y_a\rangle &=&
{\rm e}^{ie{n_{\gamma}}cL} |c,x_1,y_1,\dots,x_a,y_a\rangle \\
\Pi_0 (x,y) |c,x_1,y_1,\dots,x_a,y_a\rangle  &=&
{\rm e}^{iec(y-x)} |c, x, y, x_1,y_1,\dots,x_a,y_a\rangle \\
\Pi_1(x,y) |c,x_1,y_1,\dots,x_a,y_a\rangle  &=&
{\rm e}^{iec(y-x)} \sum_i \delta (y-x_i)
|c,x_1,y_1,\dots,x,y_i,\dots,x_a,y_a\rangle \\
\Pi_2(x,y) |c,x_1,y_1,\dots,x_a,y_a\rangle  &=&
- {\rm e}^{iec(y-x)} \sum_i \delta (x-y_i)
|c,x_1,y_1,\dots,x_i,y,\dots,x_a,y_a\rangle \\
E(z) |c,x_1,y_1,\dots,x_a,y_a\rangle &=&
\frac{i\hbar}{L} \frac{\partial{}}{\partial c}
|c,x_1,y_1,\dots,x_a,y_a\rangle \nonumber\\
&+&  e\hbar \left[ \sum_k \theta(x_k,y_k,z)
- \sum_k \frac{(y_k-x_k)}{L}\right] |c,x_1,y_1,\dots,x_a,y_a\rangle
\label{EREP}
\end{eqnarray}

Translation of the boundary conditions in this representation yields
\begin{eqnarray}
|c,\dots,x_i,y_i+pL\dots\rangle
&=& {\rm e}^{ip\pi}\sum_n {\rm
e}^{iec\left[(n+p)L+\sum_k(y_k-x_k)\right]}
|n+p,\dots,x_k,y_k\dots\rangle \nonumber \\
&=& {\rm e}^{ip\pi}|c,\dots,x_i,y_i\dots\rangle.\,
\end{eqnarray}

The Hamiltonian can be rewritten now as
\begin{eqnarray}
&H& \Psi(c,x_1,y_1,\dots,x_a,y_a) = \nonumber \\
& & \frac{e^2\hbar^2}{2}
\left\{ \int_0^L dx
\left[ \frac{i}{eL} \frac{\partial~}{\partial c}
 + \sum_k \left(\theta(x_k,y_k;x)-\frac{(y_k-x_k)}{L}\right)
\right]^2
\right\} \Psi(c,x_1,y_1,\dots,x_a,y_a) \nonumber\\
&-& i\hbar \lim_{\epsilon\rightarrow 0}
\frac{\partial~}{\partial\epsilon}
\sum_i \left\{ {\rm e}^{iec\epsilon}
            \left[\Psi(c,\dots,x_i,y_i-\epsilon,\dots)
                    + \Psi(c,\dots,x_i+\epsilon,y_i,\dots)\right]
             \right\}\,.
\label{zmodeH}
\end{eqnarray}
Now it is convenient to disentangle the term
\begin{equation}
X=  \int_0^L dx
\left[ \frac{i}{eL} \frac{\partial~}{\partial c}
 + \sum_k \left(\theta(x_k,y_k;x)-\frac{(y_k-x_k)}{L}\right)
\right]^2 \equiv X_1 + X_2,
\label{SQR}
 \end{equation}
in Eq. (\ref{zmodeH}), where we recall that
\begin{equation}
{\bar \theta}(x_k,y_k;x)\equiv \theta(x_k,y_k;x)-  \frac{y_k-x_k}{L}
=\frac{1}{L}  \sum_{p\ne 0} \frac{1}{ip}
                   \left( e^{ip y_k} - e^{ip x_k}
                   \right) e^{-ipx}\, ,
\end{equation}
 according to Eq.(\ref{THETA}). Since the summation in the  
above equation
is  over $ p\neq 0$, the integration  of the crossed term in the  
square of Eq.
(\ref{SQR}) is  zero. The integration  of the first square term  is  
immediate,
leading to
\begin{equation}
X_1= -\frac{1}{e^2 L} \frac{\partial^2}{\partial c ^2}.
\end{equation}
The integration of the second square term  reduces to
\begin{eqnarray}
X_2&=& \int_0^L dx \sum_{k,l} {\bar \theta}(x_k,y_k;x){\bar
\theta}(x_l,y_l;x)\nonumber \\
&= &L \sum_{k,l} \left(  V(y_k-y_l) + V(x_k-x_l) - V(y_k-x_l) -  
V(x_k-y_l)
\right),
\end{eqnarray}
where
\begin{equation}
V(z) = \sum_{n\neq 0} \frac{1}{ 4 \pi^2 n^2} e^{ \frac{2 \pi i  n   
}{L} z}.
\label{POTEN}
\end{equation}

The main conclusion of the above  calculation is the fact the the  
solutions of
 the whole problem
satisfy the separability condition
\begin{equation}
\Psi(c, \dots x_{ I },y_{ I}, \dots) = \Phi(c)  \
\Theta( \dots, x_{I},y_{I},\dots)\,. 
\end{equation}

The external field analysis performed in Section  IV  corresponds to
neglecting the quadratic (electric field) term  in (\ref{zmodeH})
with respect
to  the fermionic contribution involving the $\epsilon\rightarrow
0$ limit.
The lowest energy state should be associated to the Dirac sea and the
corresponding
energy is  $E(c)= -\frac{\hbar \pi}{6L} + \frac{\hbar L}{2\pi} e^2 c^2$,
$-\frac{\pi}{eL}\leq c\leq \frac{\pi}{eL}$.

The next step is to consider the zero-mode sector of the theory,  which
consists in
taking
$E(z) \rightarrow E_0= {1\over L} \int dz  E(z) $ in the Hamiltonian
(\ref{zmodeH}). It is important to remark that the separability of Eq.  
(\ref{zmodeH}) in
the form
 $\Psi(c,...x_I,y_I...)=\Phi(c)\Theta(...x_I,y_I...)$,
implies that, in the compact case, the zero mode
spectrum is a  part of the exact spectrum of the  full 
Schwinger model,
in complete analogy with  the noncompact case.

From the
general expression (\ref{EREP}), and recalling Eq. (\ref{THETA}) we
readily
verify that
\begin{equation}
E_0= {i\hbar\over L}{\partial \over \partial c},
\label{E0}
\end{equation}
in this representation. It is a general property that $E_0 |
0\rangle_D=0$,
which means that
$\Theta_D= \langle c,\dots,x_I,y_I \dots | 0\rangle_D$ is independent
of the
coordinate  $c$.

 Now,  the contribution of the
zero mode to the  exact solution of the problem, corresponds to the  
choice
\begin{equation}
\Psi(c, \dots x_{ I },y_{ I}, \dots) = \Phi(c)  \
\Theta_{D}( \dots, x_{I},y_{I},\dots)\,. 
\end{equation}
Recalling that the fermionic part of the Hamiltonian acts like a
derivative,
the zero mode contribution $\Phi(c)$ fulfills the equation
\begin{equation}
\left[-\frac{\hbar^2}{2L} \frac{\partial^2~}{\partial c^2}
      + \frac{\hbar L}{2\pi} {e^2}{c^2} \right] \Phi(c) = E_1 \Phi(c) \,
\label{zeromode}
\end{equation}
where the constant $-\frac{\hbar\pi}{6L}$ coming from the Dirac sea energy
$E(c)$ was reabsorbed in $E_1$.

The inner product and the boundary conditions
for the zero mode contributions are:
\begin{eqnarray}
\langle\phi|\psi\rangle &=& \int_{-\frac{\pi}{eL}}^{\frac{\pi}{eL}} dc\
                     \phi^{\ast}(c)\psi(c), \nonumber \\
\phi(-\frac{\pi}{eL})&=&
\phi(\frac{\pi}{eL}), \quad
\phi'(-\frac{\pi}{eL})=
\phi'(\frac{\pi}{eL}), \quad
-\frac{\pi}{eL} \leq c \leq \frac{\pi}{eL},
\end{eqnarray}
in such a way  that the electric field and the
Hamiltonian are hermitian. Here the prime means the derivative of the
corresponding
function  with respect to the argument.

The solution to the eigenvalue equation
(\ref{zeromode}) can be expressed in terms of cylindrical parabolic
functions
\cite{Abramowitz} upon the change of variables
\begin{eqnarray}
&&x:=\sqrt{ \frac{2eL}{\sqrt{\pi\hbar}} } c, \quad
E_1=-\sqrt{\frac{\hbar^3}{\pi}}e a,
\nonumber \\
&&\psi(c)\rightarrow
y(x), \quad
 \quad - x_M \leq x\leq x_M, \quad
x_M\equiv \sqrt{ \frac{2}{eL}\sqrt{\frac{\pi^3}{\hbar}} },
\label{CHANGVAR}
\end{eqnarray}
where $x$ and $a$ are dimensionless quantities. This yields the equation
\begin{equation}
y''-\left(\frac{1}{4} x^2 + a\right)y =0.
\label{ZMODEQ}
\end{equation}
The periodic  boundary conditions on  $y(x)$ are
\begin{equation}\label{PBCG}
y(x_M)=y(-x_M), \quad y'(x_M)=y'(-x_M).
\end{equation}
The general solution of Eq.(\ref{ZMODEQ}) is
\begin{equation}
y(x)=A \ {\rm e}^{-\frac{x^2}{4}}\,
{\rm M}\left(\frac{a}{2}+\frac{1}{4},\frac{1}{2}, \frac{x^2}{2}\right)
+ B \ x \ {\rm e}^{-\frac{x^2}{4}}\,
{\rm M}\left(\frac{a}{2}+\frac{3}{4},\frac{3}{2},
\frac{x^2}{2}\right) \,,
\end{equation}
where  ${\rm M}(A,B,z)$ is the confluent
hypergeometric function. It will be convenient to introduce the new
 label
\begin{equation}\label{NEWVAR}
{\tilde  l} := {2 \pi^{3/2}
\over {x_M}^2}= eL \hbar^{1/2}:= { l}  \pi^{3/2}.
\end{equation}

Now we separately discuss the even and odd solutions:

(i) Even solutions:  in this case the periodic boundary conditions
(\ref{PBCG})
 are
automatic on $y$ and
imply
\begin{equation}
y'(x_M)=y'(-x_M)=0.
\end{equation}
This  eigenvalue condition for Eq. (\ref{ZMODEQ})
will determine the
energy $E_1(a)$ as a function of $l$   . The above condition can
be written as
\begin{equation}
{\rm M}\left(\frac{a}{2}+\frac{1}{4},\frac{1}{2},
\frac{{x_M}^2}{2}\right)=
(2a +1) {\rm M}\left(\frac{a}{2}+\frac{5}{4},\frac{3}{2},
\frac{{x_M}^2}{2}\right)
\end{equation}
and it defines the function $a = a (l)$. This function can only be
determined
numerically for arbitrary $l$ and it is shown in Fig. 1.

The novel properties are:
\begin{enumerate}
\item $\lim_{L\rightarrow 0} a(l)= a(0) = -\frac{1}{2} -2n, \quad
n=0,1,2,3\dots $  which
reproduces the even  subset of  the  standard
U(1) non compact case (i.e. $-\infty\leq c\leq\infty)$. We will label the
even functions
$a(l)$ by the integer $2n$: $a_{2n}(l)$. In Fig. 1. we use the
notation $ En$,
i.e. $E0, E1, E2, \dots, $ for the corresponding solutions.

\item From  the numerical calculation  we find  that  $a_0$ is
monotonously
increasing
and  also that $\lim_{L\rightarrow \infty} a_0(l)= a_0(\infty) =
0$. This
last property
is consistent with the fact that if $a$ remains finite when $l\rightarrow
\infty$, then $a=0$.

\item  The behavior of $a_{2n}( l)$ for ${
l}\rightarrow0$ is
\begin{equation}
a_{2n}( l)=-{1\over 2} - 2n + { 2 {\rm e}^{ -{1\over
l}} \over
{ l}{}^{(2n + {1\over 2})}  \ n! \  \Gamma(n+{1\over2})}.
\end{equation}
{}From the above equation we conclude that  $a_{2n}'(
l)|_{ l =
0}=0$ and also that
$a_{2n}( l)$ is an increasing function near $ l =0$.

\item The behavior for  negative $a_{2n}$, with large absolute value (
$|a_{2n}|>>1$),  is given by
\begin{equation}
a_{2n}( l)=-{\pi^2\over 2} n^2  l, \quad n= 1, 2, 3 \dots,
\end{equation}
which can be readily observed in Fig.1.
\end{enumerate}

(ii) Odd solutions:  in this case the periodic boundary conditions
(\ref{PBCG})
 are
automatic on $y'$ and
imply
\begin{equation}
y(x_M)=y(-x_M)=0.
\end{equation}

This  eigenvalue condition is
\begin{equation}
{\rm M}\left(\frac{a}{2}+\frac{3}{4},\frac{3}{2},
\frac{{x_M}^2}{2}\right)=0
\end{equation}
and it defines the odd sector of the  function $a = a (l)$. Again,  this
function can only be determined numerically for arbitrary $l$ and
it is shown
in Fig. 1. The  properties are:

\begin{enumerate}
\item $\lim_{L\rightarrow 0} a(l)= a(0) = -\frac{1}{2} -(2n-1), \quad
n=1,2,3\dots $,   which
reproduces the odd  subset of  the  standard
U(1) non compact case (i.e. $-\infty\leq c\leq\infty)$. We will label the
functions
$a(l)$ by the integer $2n-1$: $a_{2n-1}(l)$. In Fig.1. we denote by
$On$, i.e.
$O1, O2, \dots, $
the corresponding  solutions.

\item From  the numerical calculation  we find  that  $a_{2n-1}(l)$ are
monotonously decreasing functions.

\item  The behavior of $a_{2n-1}( l)$ for $ l\rightarrow0$ is
\begin{equation}
a_{2n-1}( l)=-{1\over 2} - (2n-1) - { 2 {\rm e}^{ -{1\over
 l}}
\over { l}{}^{(2n - {1\over 2})}  \  (n-1)! \ \Gamma(n+{1\over2})}.
\end{equation}
{}From the above equation we conclude that  $a_{2n-1}'(
l)|_{ l =
0}=0$ and also that
$a_{2n-1}( l)$ is a decreasing  function near $ l =0$.

\item The behavior for  negative $a_{2n-1}$ with large absolute value (
$|a_{2n-1}|>>8$) is given by
\begin{equation}
a_{2n-1}( l)=-{\pi^2\over 2} n^2  l, \quad n= 1, 2, 3 \dots,
\end{equation}
which again  can be readily observed in the figure.
\end{enumerate}

Let us notice  that  the asymptotic behavior  (large ${ l}$)
of  $a_{2n}$
and $a_{2n-1}$
coincide for $n=1,2, \dots$

 Finally, we notice that all the states in the compact case are
invariant under large gauge transformations. In particular, that is the
case for the vacuum, and consequently the $\theta$-dependence does not
appear.

 \section{Concluding remarks}

In this paper we have analysed the two dimensional compact
electrodynamics using the loop representation. 
 The loop variables introduced here adapt naturally to the study of the compact version
of the Schwinger model in the continuum, which, up to our knowledge, has not been previously discussed in the literature. These variables provide gauge invariant and intrinsically regularized non-local fields to describe the model. In terms of them, the quantum commutator algebra is constructed in an  unambiguous manner. In particular, the current-current and electric field-current commutators are directly recovered. The choice of our basic non-local variables is such that the algebra involved in  the anomalous commutators is already  realized in terms of the classical Poisson brackets . We also obtain full operator expressions for the divergences of the charged and axial currents, which still need to be analyzed in the  Hilbert space of the problem. This work also reports  the spectrum of the zero mode sector of the compact Schwinger model, which is an exact piece of the full spectrum, due to the separability of the Schroedinger
equation. The zero mode energy turns out  to be completely different from the standard equally-spaced oscillator spectrum of the non-compact case. 
Even though bosonization seems to be still present, the zero mode
spectrum suggests that the compact system will not behave as a free massive
scalar field. 

The invariance of the
theory under large gauge transformations insures the uniqueness of the
vacuum. The axial anomaly is still present, however as it was noticed 
by Jackiw
\cite{Ja}, the presence of an axial anomaly is not necessarily related
with a non trivial topological structure. Further studies are required
to determine  that, either the chiral charge may be redefined in a gauge
invariant way,  or the conservation of the axial current cannot be
restored in the compact case.
Our main  obstacle to proceed with the solution of the full  compact model is to obtain a complete set of solutions of  the Schroedinger equation  arising from the Hamiltonian (\ref{zmodeH}),  which  is of the many-body type, with a linear kinetic energy term and pairwise interactions given by the  non-trivial potential (\ref{POTEN}).There remains also the task of  giving   a complete discussion
of the structure of the Hilbert space. These
problems are now under investigation.
 
Again,  we emphasize that the choice between the compact or  
noncompact version
of  any
gauge theory  is just a mater of  convenience,  which should be  
ultimately
decided in terms
of the experimental consequences  of each model.  In this spirit, what is
really missing in the vast literature regarding the Schwinger  
model is, up
to our knowledge,  a study of  its compact version using one's  
 favorite
method of solution. The main difference between the compact and the  
noncompact
version  is expected to   appear
in the boundary conditions satisfied by the corresponding wave  
functions, as
opposed to the form of the (functional) differential equations  
involved, which
should be the same in both cases. This is in complete analogy with   
the case of
a free particle in a line (noncompact case)  compared with the free
one-dimensional rotator (compact case), both of which are governed by  
 the same
Schroedinger equation.

An  alternative route in understanding the compact Schwinger model   
is been
pursued in Ref.\cite{LMUV},
following the
method of Ref.\cite{IsoMurayama}.
Some preliminary results are the following. Since the Gauss law  
still implies
that the wave function is independent of the excited modes of the vector
potential in the Weyl gauge, there are no boundary condition  
modifications
related to  these variables arising from the compactification. The  wave
function depends only upon the zero mode of the vector potential  
and the full
Hamiltonian is  still  separable into zero plus excited modes with  
the same
Schroedinger equations
as in the noncompact case. The zero mode
wave function  satisfies  new boundary conditions, imposed by the
compactification of the zero mode vector potential, leading to the same
non-linearly spaced spectrum derived in the present
work. This piece of the spectrum cannot be interpreted as a collection of particles  
with mass  $
\frac{e}{\sqrt  \pi}$ at zero momentum.  The study  of the complete spectrum and the  full Hilbert space of the compact
Schwinger model, using this approach,  is also in progress.\cite{LMUV}

\acknowledgments
Partial support from CONACyT grant 3141-PE to HAMT,  LFU and JDV is
acknowledged.
HAMT would also like to thank
CLAF-M\'exico, the  organizers
of ``Quantum Gravity in the Southern Cone"
and ICTP-Trieste for  hospitality and financial support at
different stages of
the present work.
HAMT and LFU are  also indebted  to  R. Gambini  for the
hospitality extended
to them at the
Departamento de F\'{\i}sica
de la Universidad de la Rep\'ublica, Uruguay. LFU and JDV  are
supported  by
the
grant DGAPA-IN100694 and DGAPA-IN100397. We thank Rom\'an Linares for verifying
some of the
calculations. RG acknowledges useful conversations with H. Fort regarding
compact electrodynamics.

\section*{Appendix A}

Using the scalar product given in Eq.(\ref{PE}), we are able to
solve for the creation-annihilation operators in terms of the fields
$\psi_1,\psi_2$
\begin{equation}
a_m=e^{{i\over \hbar}\epsilon_m t}(\phi_m,\ \psi_1),\quad
a_m^\dagger=e^{-{i\over \hbar}\epsilon_m t}(\phi_m^*,\ \psi_1^*),
\end{equation}
\begin{equation}
b_m=e^{{i\over \hbar}\epsilon_m t}(\phi_m^*,\ \psi_2^*),\quad
b_m^\dagger=e^{-{i\over \hbar}\epsilon_m t}(\phi_m,\ \psi_2).
\end{equation}
In this way, all bilinear expressions containing creation-annihilation
operators can be written in terms of the gauge-invariant operators
$\Pi(x,y)$
defined at $t=0$.

A direct calculation shows that
\begin{eqnarray}
\Pi_0^{mn}&=& a_m^\dagger b_n^\dagger, \\
\Pi_1^{mn}&=& \,a_m^\dagger a_n, \\
\Pi_2^{mn}&=& -b_n^\dagger b_m, \\
\Pi_3^{mn}&=& b_m a_n\;.
\end{eqnarray}
The fermionic character of the operators involved imply the following
important symmetry properties of the above operators
\begin{equation}
\Pi_a^{mn}\Pi_a^{mk}=0=\Pi_a^{mn}\Pi_a^{km},\quad a=0,3,
\end{equation}
which is reflected through $\Pi_a(x,y)\Pi_a(x,z)=0=\Pi_a(x,y)\Pi_a(z,y)$
in coordinate space. Another important related symmetry of the external
field is $U(u,v)U(x,y)=U(u,y)U(x,v)$, which results because of the
abelian character of the $U(1)$ connection.

\newpage
\centerline{\psfig{figure=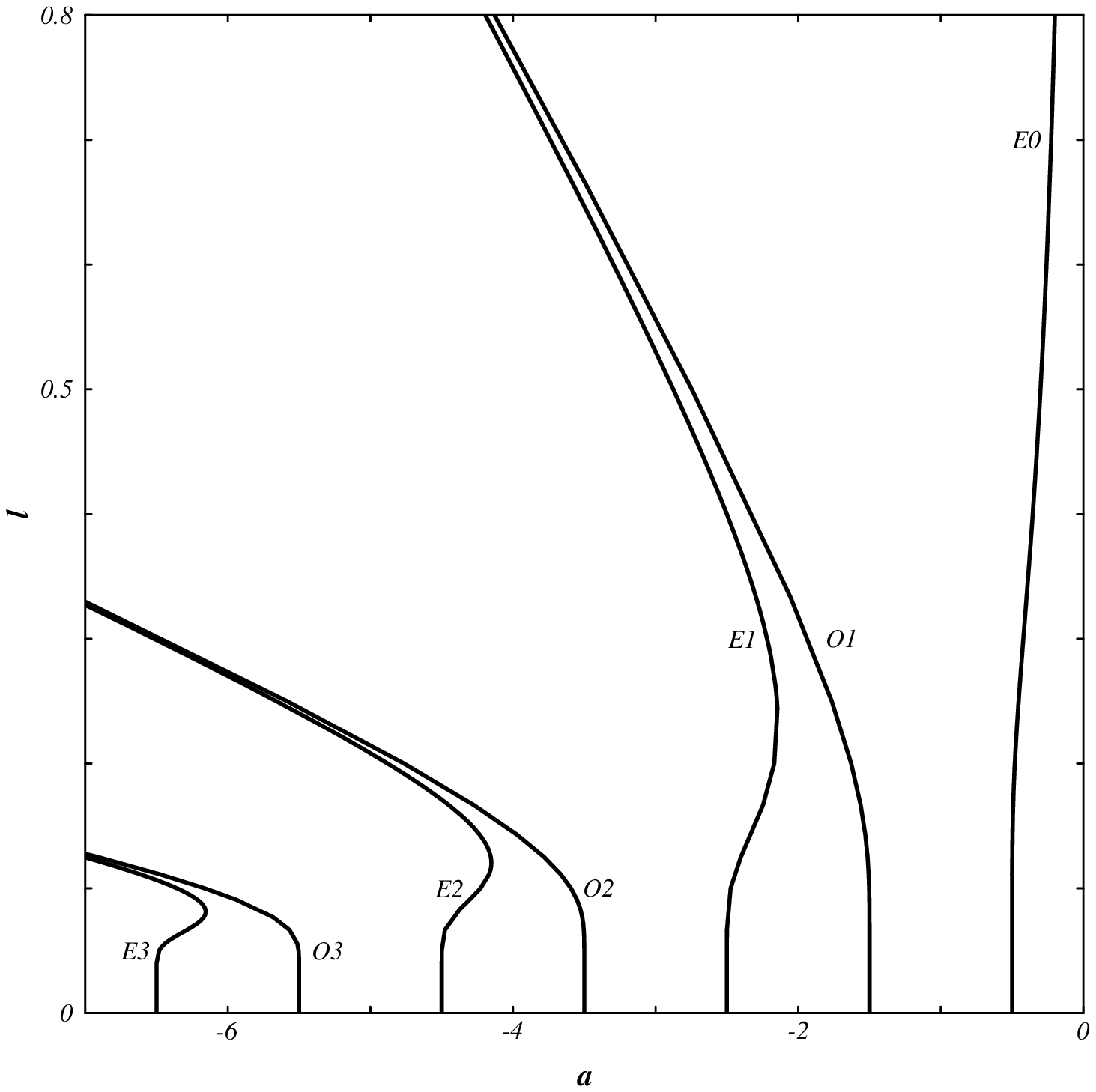,height=23cm}}
\vskip-7cm
{FIG.1. For a given  value of $l$, the above curves provide the
numerical solution for the parameter $a(l)$,  which is related to the
zero-mode spectrum  of the system  through $E_1=-(\hbar^3/\pi  
)^{1/2} e a(l)$.
The even (odd) solutions
are labeled
by  $En (On), n=0,1,2, \dots, $  and correspond to the solutions
$a_{2n}(l)$ $(
a_{2n-1}(l))$ of  Section VII.
}


\begin{references}

\bibitem {Schwinger}J. Schwinger, Phys. Rev. {\bf 125}, 397 (1962);
         {\em ibid} {\bf 128},
          2425 (1962).
\bibitem {general} L.S. Brown, Nuovo Cim. {\bf XXIX} 3727 (1963);
       J.H. Lowenstein and
         J.A. Swieca, Ann. Phys. {\bf 68} 172 (1971);
        E. Abdalla and M.C.A. Abdalla, {\em Non-perturbative methods
    in 2 dimensional quantum field theory}
  (World Scientific, Singapore, 1991).
\bibitem{ADAMS} For a recent review of the Schwinger model see for
example C. Adams, {\it  Anomaly and Topological aspects of two-dimensional
quantum electrodynamics}, Dissertation, Universitat Wien, october 1993.
\bibitem {Manton} N.S. Manton, Ann. Phys. {\bf 159}, 220 (1985).
\bibitem{HetHo} J.E. Hetrick and Y. Hosotani, Phys. Rev. {\b D38},  
2621 (1988).
\bibitem {Shifman} M.A. Shifman, Phys. Rep. {\bf 209}, 341 (1991).
\bibitem{Link} R. Link,   Phys. Rev. {\b D42}, 2103 (1990).
\bibitem{IsoMurayama} S. Iso and H. Murayama, Prog. Theo. Phys.
{\bf 84},142 (1990).

\bibitem{HallinLiljenberg} J. Hallin and P. Liljenberg,  Phys.
Rev. {\bf D54},
1723 (1996);  J. Hallin, QED$_{1+1}$ by Dirac Quantization,
Preprint
G\"oteborg, ITP 93-8, hep-th/9304101, May 1993.

\bibitem{G-P} R. Gambini and J. Pullin, {\it Loops, knots, gauge
theories and
quantum gravity} (Cambridge: Gambridge Univ. Press 1996).

\bibitem {RovelliSmolin} C. Rovelli and L. Smolin, Phys. Rev. Lett.
{\bf 61},
         1155 (1988);
         Nucl. Phys. {\bf B133}, 80 (1990).

\bibitem{Gambini} R. Gambini, Phys. Lett.  {\bf B235 }, 180 (1991).

\bibitem {GambiniTrias} R. Gambini and A. Trias, Phys. Rev. D {\bf
23}, 553 (1981); Nucl. Phys. {\bf B256}, 479 (1986).

\bibitem {FortGambini} H. Fort and R. Gambini, Phys. Rev. D {\bf 44},
        1257 (1991).

\bibitem {HugoRovelli} H.A. Morales-T\'ecotl and C. Rovelli, Phys.
Rev. Lett.
         {\bf 72}, 3642 (1994); Nucl. Phys. {\bf B451}, 325 (1995).

\bibitem{Po} A.M. Polyakov,{\it Gauge Fields and Strings}
(New York: Harwood 1987).

\bibitem{COMPACT}  F. Palumbo, Phys. Letts. {\bf  B225}(1989)407.; {\it ibid.} {\bf  B335}(1994)215.
\bibitem{FA} J.M.Aroca,  H.Fort and Gonzalo Alvarez , {\it Finite lattice hamiltonian computations in the P-representation: the Schwinger model}, preprint hep-lat/9711049.
\bibitem{F}  H.Fort,  {\it  The worldsheet formulation as an alternative method for
simulating dynamical fermions},  preprint  hep-lat/9710082.
\bibitem{Poly96} A.M. Polyakov, Nucl. Phys. {\bf B486}, 23 (1997).

\bibitem{GAMB5}  H.Fort and R. Gambini,  {\it The U(1) and strong CP problem from
the loop formulation perspective},  preprint hep-th/9711174.

\bibitem{SCHCOM} J. Schwinger, Phys. Rev. Letts. {\bf 3}, 296 (1959).

\bibitem{Abramowitz} M. Abramowitz and I. A. Stegun, {\it Handbook of
Mathematical Functions} (New York: Dover 1965).

\bibitem{Ja} R. Jackiw Topological investigations on quantized gauge
theories. {\it Relativit\'e Groupes et Topologie II, Les Houches 1983}
B. DeWitt and R. Stora, Editors (North-Holland, Amsterdam 1986)

\bibitem{LMUV} R. Linares, H. A. Morales, L. F. Urrutia and
J. D. Vergara, work in progress.


\end{references}
\end{document}